\documentclass[article,amsmath,amssymb]{revtex4}

\usepackage{graphicx}% Include figure files
\usepackage{dcolumn}% Align table columns on decimal point
\usepackage{bm}% bold math
\usepackage{amsmath,amssymb}
\begin{document}

\title{ Vibrational spectrum of the H$_{5}^+$ molecule \\ using quantum
Monte Carlo}
\author{Washington Barbosa da Silva
, Luiz Roncaratti, Geraldo Magela e Silva, and Paulo Hora Acioli}
\affiliation{
Instituto de F\'{\i}sica, Universidade de Bras\'{\i}lia,
CP 04455, 70919-970  Bras\'{\i}lia - DF, Brazil
}

\date{ }

\begin{abstract}
In this article we present a caracterization of the vibrational spectrum of the
H$_{5}^+$ molecule using the correlation function quantum Monte Carlo (CFQMC)
method and a genetic algorithm study of the topology of the potential energy
surface used in this work. The vibrational modes associated with the
H$_3^{+}$-H$_2$ torsion and stretching posses very flat minima. As a
consequence the fundamental frequencies corresponding to these modes are
poorly described in the  harmonic approximation.
The vibrational frequencies obtained in this work are in good agreement with
the available experimental data as well as other computational
methods found in literature. In our genetic algorithm study of the potential
energy surface using cartesian coordinates we have found some unexpected minima.
A careful analysis shows that some of these minima are described by the same
curviliniar coordinates in which the potential is described. However,
they represent nonequivalent molecular geometries.
\end{abstract}
\maketitle

\section{Introduction}
The presence of a molecular or atomic ion in a H$_2$ atmosphere results in the formation of hydrogen ionic clusters X$^+$(H$_2$)$_n$. Properties of these clusters, such as solvation-shell distributions \cite{a4,MGM01}, binding energies \cite{YGRS87,a5,RAPG01}, and vibrational spectrum \cite{RAPG01,MM03}, have been theoretically investigated for a large variety of ionic cores X$^+$.

The H$_2$ units are connected to the core mainly by means of monopole/induced-dipole interactions, although there is some charge transfer from the H$_2$ units in the first solvation shell to the positive center. As a consequence of this kind of interaction, binding energies lie on the range of 3-10 kcal/mol \cite{PJMG98,RSJ99,a0} that is not enough to destroy the molecular identity of each H$_2$ unit. On the other hand, these binding energies are enough large to make the hydrogen clusters potentially useful in storage of hydrogen fuel, by taking advantage of the cluster formation in the positive-charged sites of adequate crystalline materials \cite{PXJK99,a7}.

The first species of the homogeneous series H$_3^+$(H$_2$)$_n$, the H$_5^+$, is an atypical hydrogen cluster \cite{a2}. Its electronic properties does not fit in those of the large members of the series \cite{MGM01,MM03}. As Ohta {\it et al.} have recently shown \cite{a2}, one main reason for such anomalous behavior is the isomerization process
\begin{equation}
H_3^+(H_2) \rightleftharpoons H_2(H^+)H_2 \rightleftharpoons (H_2)H_3^+ ,
\end{equation}
which take places even for very low temperatures.

Since its detection in 1962 \cite{DT62}, many experimental studies of dissociation energies
and thermochemical properties of the H$_{n}^+$ clusters have been performed
\cite{YGRS87,OYL88,NMT92,FBM01}.
For the purpose of the present work, the investigations performed by
Okumura $\it{et \hspace{0.1cm}al}$ \cite{OYL88} using infrared vibrational predissociation spectroscopy
are particularly relevant.
They observed the broad and structureless vibrational bands corresponding to the
H$_2$ and the H$_{3}^+$ units into the (H$_2$)$_{n}$ ionic clusters for n = 1-6 and analyzed
the dependence of the shift of these vibrations with the size of the cluster. A similar approach
was latter used by Bae \cite{B91}.

Numerous $\it{ab \hspace{0.1cm}initio}$ calculations have been carried out
\cite{a4,MGM01,YGRS87,a5,RAPG01,MM03,a7,a2,NMT92,AR75,YGS83,IDMK97,MBNH91,a1,a3,a6,WVO94}
aiming to determine their equilibrium structures,
low-lying stationary points and dissociation energies. A very detailed
$\it{ab \hspace{0.1cm}initio}$ investigation of the H$_{5}^+$ cluster was carried
out by Yamaguchi $\it{et \hspace{0.1cm}al}$ \cite{YGRS87}. They revealed a quite
complicated structure of the interaction potential: ten distinct stationary points
were located.
Their high-level calculations predicted a C$_{2v}$ structure as the global minimum.
\v{S}pirko {\em et al.}
studied the potential energy surface of H$_{5}^+$ and its corresponding
vibrational spectrum using the multi-reference configuration interaction  (MR-CI)
methodology \cite{S93,WVO94,S97}.
Besides these works
there are only a few theoretical papers dedicated to the infrared spectrum
of H$_{5}^+$ and they are all retricted to the harmonic frequencies
\cite{YGRS87,RAPG01,a7,MBNH91,a1}. A recent work by Barbatti and
Chaer Nascimento  examined in some depth the vibrational spectra of
H$_{2n+1}^{+}$ using
the VSCF methodology \cite{MM03}. As mentioned above, in addition to the scarce
theoretical data
only the experimental results of Okumura \cite{OYL88} and Bae \cite{B91} are available for the vibrational spectrum of this system. %<<<<
Therefore a more complete caracterization of this system is warranted. The focus of this work is to provide a detailed description of
the vibrational spectrum of H$_{5}^+$.

A full description of the vibrational spectrum of H$_{5}^+$ needs to
take into account nine degrees of freedom. The computational effort in the
traditional variational methods that rely on integration using normal quadratures
is prohibitive. One of the alternatives to overcome such shortcoming is the use
of Monte Carlo methods to calculate the multidimensional integrals which appear
in time independent problems, \cite{CB88,BCL90,APH99,ACP99,REV00,WP01}
which do not present the
same limitations of the variational methods. This method has been applied
successfully for the computation of the vibrational and ro-vibrational spectra
of triatomic and tetratomic molecules
\cite{APH99,ACP99,REV00,WP01,PLF00,FVA99,FVN99,FVP98,WEPR03}.
In this work we used the CFQMC methodology to obtain the vibrational states of
the H$_{5}^+$ molecule. We also study the PES used in this work using the
genetic algorithm\cite{genetico1,genetico2,genetico3, genetico4, genetico5, genetico6}.
This article is organized as follows. A short description of
CFQMC methodology and of the genetic algorithm is given in  the next section.
Sec. \ref{results} is devoted to the presentation of the results, and in sec.
\ref{conclusion} the discussion and concluding remarks are presented.

\section{Methodology}
\subsection{Correlation Function Quantum Monte Carlo}
\label{cfqmcm}

We start from the Born-Oppenheimer nuclear Hamiltonian of a molecular
system (in the center-of-mass reference frame)
\begin{equation}\label{2}
H = -\sum_{i=1}^{N-1}\frac{\hbar^2}{2\mu_i}\nabla_{i}^{2} + V({\bf R}) ,
\end{equation}
where {\bf R} is the vector of the coordinates of all particles of the system,
$V({\bf R})$ is the potential energy of the nuclei. The vibrational energy levels
of the system will be then computed by solving the eigenvalue problem:
\begin{equation}\label{1}
H\Phi_i({\bf R}) = E_i\Phi_i({\bf R}),
\end{equation}
where $E_i$ and $\Phi_i({\bf R})$ are the eigenvalues and eigenvectors of $H$.
Given a trial basis set of {\it m} known functions $f_{i}({\bf R})$ we can define
the following generalized eigenvalue problem \begin{equation}\label{4}
\sum_{j=1}^{m}[H_{ij} -\Lambda_k N_{ij}]
d_{kj}=0\; ,
\end{equation}
where $d_k$  is  the  $k^{th}$  eigenvector and  $\Lambda_k$  its  associated
eigenvalue, and
\begin{eqnarray}\label{3}
N_{ij} = \int d{\bf R}f_{i}({\bf R})
f_{j} ({\bf R}),  \nonumber \\
H_{ij} = \int d{\bf R}f_{i}({\bf R})
Hf_{j}({\bf R})  \; ,
\label{eq4}
\end{eqnarray}
are the overlap and Hamiltonian matrix elements associated with the basis set.
The matrix elements defined in Eq.(\ref{eq4})  are evaluated using Monte
Carlo integration techniques. The only difference of this method and the
traditional variational methods is the use of Monte Carlo to compute the
integrals. Of course an accurate spectra will depend on the quality of the basis
set $\{f_{i}({\bf R})\}$. In this work we use the basis set of reference
\cite{APH99}. Namely, for the ground state
\begin{equation}\label{8}
\psi_0 = exp(\sum_{\nu\mu} \Delta S_\nu A_{\nu\mu}\Delta S_{\mu})
\end{equation}
with $\Delta S_{\nu} = S_\nu - S^0_{\nu}$, and $S_{\nu} = |r_i - r_j|$ the
distance between atoms $i$ and  $j$, and $S^0_{\nu} = |r_i - r_j|^0$ the
equilibrium distance between atoms $i$ and  $j$. The variational parameters
$A_{\nu\mu}$ are optimized in order to minimize the variational energy or its
variance.  The trial functions for the excited states are given by
\begin{equation}
f_{i} = \psi_0\prod_{\nu} (\Delta S_{\nu})^{n_\nu(i)}.
\end{equation}
This basis set was applied successfully in the study of vibrational spectra of 2-,
3- and 4-atom molecules
\cite{CB88,APH99,ACP99,REV00,WP01,PLF00,FVA99,FVN99,FVP98,WEPR03}.
An improvement to this methodology is
the use of the diffusion Monte Carlo techniques to ``project" out the excited state
spectrum. We restricted ourselves to the variational implementation of the method in
order to have a comparison in the same
footing of this technique with the VSCF, which is also a variational based method.

Another important aspect related to the accuracy of the CFQMC is the quality of the
potential energy surface (PES) describing the motion of the nuclei. In this work
the PES of \v{S}pirko {\em et al.} \cite{WVO94} fitted to MR-CI
all-electron computations at 110 different configurations of the nuclei was used.
A set of curvilinear coordinates was used to facilitate the computation of the
vibrational spectrum .  The purpose of the CFQMC calculations is twofold. First,
to show that the method can be applied to compute the spectrum of a 5-atom molecule.
Second, to determine the accuracy of the PES of ref. \cite{WVO94}.

\subsection{Genetic Algorithm}
\label{vscfm}

In our genetic algorithm the population is coded in a binary discrete cube named $\bf{A}$, with $l\times
m\times n$ bits. The elements of $\bf{A}$, $a_{ijk}$, are either 0 or 1, with $i,j,k$ integers numbers $1\leq
i\leq l$, $1\leq j \leq m$, $1\leq k \leq n$. The label $i$ refers to the component $i$ of the gene $j$ of
the individual $k$. Therefore, $\bf{A}$ represents a population of $n$ individuals, each one of them have a
genetic code with $m$ genes. Each gene is a binary string with $l$ bits.

The genetic code of the individual $k$ is given by
\begin{equation}
[\overline{\bf{a}}]_k=[\overline{a}_{1k},\overline{a}_{2k},...,\overline{a}_{mk}],\nonumber
\end{equation}
were
\begin{equation}
\overline{a}_{jk}=\sum_{i=1}^{l}2^{i-1}a_{ijk} \label{aijk}
\end{equation}
is a integer number composed with the binary string $a_{1jk}a_{2jk}..a_{ijk}..a_{ljk}$. It is defined on the
interval $[0,2^{l}-1]$. To define the real search space for each parameter, we transform
\begin{equation}
\overline{a}_{jk} \rightarrow {a}_{jk} \equiv
\frac{(a^{max}_{j}-a^{min}_{j})}{2^{l}-1}\overline{a}_{jk}+a^{min}_{j}\label{ajk}
\end{equation}
where ${a}_{jk}$ is a real number defined on the interval $\delta_j=[a^{min}_{j},a^{max}_{j}]$.

We define the phenotype of the individual $k$, $V_{k}\equiv V([\bf{a}]_{k})$ where
$[\bf{a}]_{k}=[a_{1k},a_{2k},...,a_{jk},...,a_{mk}]$
is a set of coefficients that characterize the individual $k$. With this we define the fitness of a phenotype
$k$ (set of coordinates)
\begin{equation}
F_k=S_{max}-S_k\nonumber
\end{equation}
where $S_k$ is the energy given by the SEP for this phenotype,
and $S_{max}$ is the worst individual in the population.

We use the most common operators: selection, recombination and mutation. The selection operator normalize the
vector $S_{k}$
\begin{equation}
P_{k}={\frac {S_{k}}{\sum S_{k}}}\label{prob}
\end{equation}
that represents the probability of each individual been selected for a recombination through a roulette
spinning. For the purpose of this work we selected $n/2$ individuals (parents) that will generate, through the
recombination operator, $n/2$ new individuals (offsprings). So, to make a new generation we joint the $n/2$ old
strings (parents) with $n/2$ new strings (offsprings) in order to maintain the population with fixed number $n$.
The recombination operator is a cross-over operator that recombine the binary string of each gene $j$ of two
random selected individuals to form two new individuals. In this work we use a two random point cross-over.

The mutation operator flip $N_{mut}$ random selected bits in a population. We choose $N_{mut}$ to make the
probability of change of a given bit equal to $0.01$ per cent. So, in a population of $l\times m\times n$ bits,
we make
\begin{equation}
 q = \frac {N_{mut}}{l\times m\times n}\label{q}
\end{equation}
where $q$ is the probability of change of one bit.

An elitist strategy is used. It consists of copying an arbitrary number $N_{el}$ of the best individual on the population
in the next generation. It warrants that this individual will not be extinguished.

We found a large number of acceptable solutions. The set of all solutions is
the definition of search space ($\Gamma$). The length of $\Gamma$ is defined  by the number $m$ of coefficients
and the length $l$ of the binary codification. Each one of the 12 coefficients (coordinates), that define the
individual $k$,
can assume $2^l$ distinct values. So, an individual in the population is only one possibility among
$2^{l\times m}$. This value defines the length of $\Gamma$. The length of $\Gamma$ describes the number of
digits that are used to express a real value $a_{jk}$ and shows the minimal difference between two
possible values of $a_{jk}$. Being each coefficient defined on an arbitrary interval $\delta_j$,
the precision of the coefficient $a_{jk}$ is

\begin{eqnarray}
\frac{a_{j}^{max}-a_{j}^{min}}{2^l}.\label{precision}
\end{eqnarray}

If we do not have any information about the order of magnitude of the $a_{jk}$ values, we must choose the
$\delta_j$'s intervals such that they cover the greatest number of values. However, after some generations, we
obtained more precise information about the order of magnitude of each coordinate $a_{jk}$. In order to improve
the performance of a standard genetic algorithm (GA), we include in our technique the concept of dynamic search
space. It
consist in the use of information of past generations to determine the length and precision of the search space
for the next generations. For the first generations, when we have few information about the magnitude
of the coordinates, we do not need many digits to represent a real number $a_{jk}$, that is, we use a low
precision codification given by a low value of $l$. In this way, we make $\Gamma$ a ``small" search space and the
GA can find the regions of acceptable solutions faster. Once found some of these regions we can redefine the
$\delta_j$'s intervals and rise the precision rising the length of binary codification $l$. After extensive
trials of the parameters values we take $m=12$, $n=100$, $q=0,01$ and $N_{el}=10$. Beside that, we always start
the GA with a random population defined in the initial intervals
$\delta_j=[a_{j}^{min},a_{j}^{max}]=[-10,10]$ and set the initial value for the length of the binary
codification $l=12$. In this way we had a search space of length $2^{l\times m}=2^{12\times 12}=2^{144}$ and the
minimal difference of two possible values of $a_{jk}$ is $\frac{10}{2^{11}}
\simeq 49 \times 10^{-4} $. After 1000
generations we redefine $l=l+4$ and $\delta_j=[a_{j}^{min},a_{j}^{max}]$ where
$a_{j}^{min}=a_{jbest}+a_{j}^{min}\times 10^{-1}$, $a_{j}^{max}=a_{jbest}+a_{j}^{max}\times 10^{-1}$ and
$a_{jbest}$ is the fitest individual in the population found along the last 1000 generations. We set 10000
generations for each run of the GA. It should be pointed out that the algorithm is very robust and works
properly with an wide range of these parameters.

\section{Results}
\label{results}
For well-behaved systems a first, usually good, description of the fundamental
vibrational frequencies is the harmonic approximation. In Table I we present
the harmonic frequencies corresponding to the PES of ref. \cite{WVO94}, used in our
CFQMC computations,
together with those obtained at MP2/6-311G(d,3p) level as the first step in the VSCF methodology, % <<<<<<<<<<<<<
the ones obtained by Prosmiti {\em et al.} using the QCISD(T)/cc-PVQZ \cite{a1},
and the CCSD(T)/aug-cc-pVTZ of Prosmiti {\em et al} \cite{RAPG01}. The results are    %<<<<<<<<<<<<<
in reasonable agreement with
each other. The largest discrepancy is observed in the normal mode corresponding to
the second skeletal motion. Whereas the value obtained using the PES of ref.
\cite{WVO94} is 1660 cm$^{-1}$, the frequencies obtained using the MP2-VSCF, the QCISD(T)/cc-PVQZ
and the CCSD(T)/aug-cc-pVTZ are 1201, 1170, and 1174 cm$^{-1}$,              %<<<<<<<<<<<<<<<<<<<<<<
respectively. This difference may be due to either the difference in the levels
of computation, or to the fitting of the PES itself. Unfortunately, in
ref. \cite{WVO94} the frequencies in the harmonic approximation are not presented
and this question may not be uniquely answered. To gauge the validity of the
harmonic approximation one must include the anharmonic effects. These effects were
considered at the CFQMC level of computation.

In Table II we consider the anharmonic effects at the CFQMC(VMC) level of
calculation. For comparison, the results of \v{S}pirko {\em et al.}  \cite{WVO94}
methodology and some available experimental results are also included.
As mentioned
above, the computation of the spectra was performed using the CFQMC method in the
variational form.  The basis set consisted of 220  basis functions and the
parameters of the wave function used in our calculations were adjusted to
minimize the energy of the ground state and the first nine excited states.
As one can see, there is a reasonable
agreement between all the computations. In particular, the results for the
8$^{th}$ and 9$^{th}$ states obtained in reference \cite{WVO94},
which correspond to the H$_{3}^{+}$ symmetric stretching and the
H$_2$ stretching motions, are in good agreement with the experimental results
of ref. \cite{OYL88}. The CFQMC results differ a little and they are closer
to the values of the frequencies of the H$_2$ and H$_{3}^{+}$
isolated molecules.
Although the CFQMC of these frequencies in the H$_{5}^{+}$ are shifted in the
right direction the shifts are much smaller than the ones predicted by the
experiment.

In FIG. \ref{fig1} we display the fundamental frequencies of the H$_2$,
H$_{3}^{+}$, and H$_{5}^{+}$ molecules in the harmonic approximation, using
the CFQMC, and the experimental results \cite{OYL88}. A similar analysis of the
MP2-VSCF results is presented in ref. \cite{MM03}. In all cases the frequency
corresponding to the symmetric stretching of H$_{3}^{+}$ is blue-shifted in the
H$_{5}^{+}$ complex and the one corresponding to the H$_2$ stretching motion is
red-shifted. However, the shifts predicted by the CFQMC are much smaller than the
ones predicted by the harmonic approximation, the experiment and the MP2-VSCF
computations. This difference may be due to the coordinates used to fit the PES
used in the CFQMC computations. This PES uses a set of curvilinear coordinates
\cite{WVO94} which are ill-defined as two completely different and non-equivalent
set of configurations of the five atoms correspond to the same set of curvilinear
coordinates. In the case of the harmonic approximation or in the original
calculations of ref. \cite{WVO94}, the computations are performed close to
the minimum and this ill-definition of the coordinates do not affect the final
frequencies. In the case of the CFQMC the integral is evaluated in the whole
space, therefore, making the results more susceptible to the choice of coordinates
used to describe the PES. Nevertheless, our calculations seem to be describing
the stretching motion of the H$_2$ and the symmetric stretching of H$_{3}^{+}$ as
independent of all the other vibrational modes.

The CFQMC results are limited by the quality of the PES but has the advantage
that one can compute the full spectrum, limited of course by the size of the basis
set. In FIG. \ref{fig2} we present the results of all the vibrational frequencies
obtained in our CFQMC up to 5000 $cm^{-1}$. The lines are broadened by 30$cm^{-1}$.
The full arrows indicate the fundamental frequencies as obtained by our CFQMC
computations, while the dotted arrows indicate the experimental results
\cite{OYL88}.
The experimental results fall in overtones of our computed results rather than
fundamental frequencies.

Because of the disagreement between the CFQMC computed frequencies of the
H$_2$ stretch
and the symmetric stretching H$_{3}^{+}$ in H$_{5}^{+}$, with the experiment and other computations we decided to  perform an additional study of the topology
of the PES used in this work. We decided to search for other minima besides the
accepted C$_2v$ minimum. This search is performed with the genetic algorithm
which has been demonstrated to be a very robust method for global minimum
search.

The inital population was completely random and after a few generations
we observed the different minima (local and global). Fig. \ref{fig3}a shows two
candidates for the global minimum. The energies of the two minima are
identical. Further analysis shows that in the curvilinear coordinates in which
the potential is defined they are identical minima.

Based on the minima of Fig. \ref{fig3}a we started a new population and run
the program for a few more generations. The minima of Fig. \ref{fig3}a are
again among the lowest energy structures. However, a new lower energy minimum
is found. This minimum is displayed on Fig. \ref{fig3}b. As one can see this
candidate for global minimum is a configuration in which the H$_2$ and the
H$_{3}^{+}$ molecules appear as not bound. This explains why the CFQMC
computations frequencies of the H$_2$ stretch and the symmetric stretching
H$_{3}^{+}$ modes are very close to their free values. The CFMQC integration
samples configuration in the whole space and they will include the structure
of Fig. \ref{fig3}b. The same does not happen with the harmonic frequencies
as they were computed near the accepted C$_2v$ global minimum.

\section{Conclusion}
\label{conclusion}

We have obtained the vibrational energies of the H$_{5}^+$ cluster in the harmonic
approximation and by using  the
correlation function quantum Monte Carlo (CFQMC) methodology.
The lowest frequencies of the H$_{5}^{+}$ are not well described in the
harmonic approximation due to the flatness of the potential energy surface
of the complex.
The only available experimental data for this system are the frequencies of the
stretching motion of the H$_{2}$ and the symmetric stretch of H$_{3}^+$ obtained
by IR predissociation experiments in the H$_{5}^{+}$ conplex. These frequencies
are red- and blue-shifted when compared to these frequencies in the free molecules.
We have seen that the shifts using CFQMC are smaller than the shifts obtained by
the harmonic approximation and the experiment.

Our genetic algorithm search of a global minimum revealed two interesting
facts about the PES of ref. \cite{WVO94}. First, their choice of  curvilinear
coordinates is ill-defined, non-equivalent configuration in cartesian
coordinate have identical curvilinear coordinates. Second, and most important,
the global minimum is a configuration in which the H$_{2}$ and H$_{3}^+$
molecules are not bound. This explains the small values of the shifts of the
frequencies of the stretching motion of the H$_{2}$ and the symmetric stretch of H$_{3}^+$ when compared to their values in the free molecules.

In conclusion, our CFQMC and genetic algorith study of the PES of ref. \cite{WVO94} shows a good agreement with
other computations and the available experimental data. The differences are explained in terms of the definition
of the curvilinear coordinates used in the definition of the PES. As a future work we are proposing to
reparametrize the PES in more appropriate coordinates with additional CFQMC computations.

\section{Acknowledgments}

This work has been supported by CNPq and CAPES through grants to the authors.

\newpage
\begin{center}
\begin{figure}
\includegraphics[0,0][424,465]{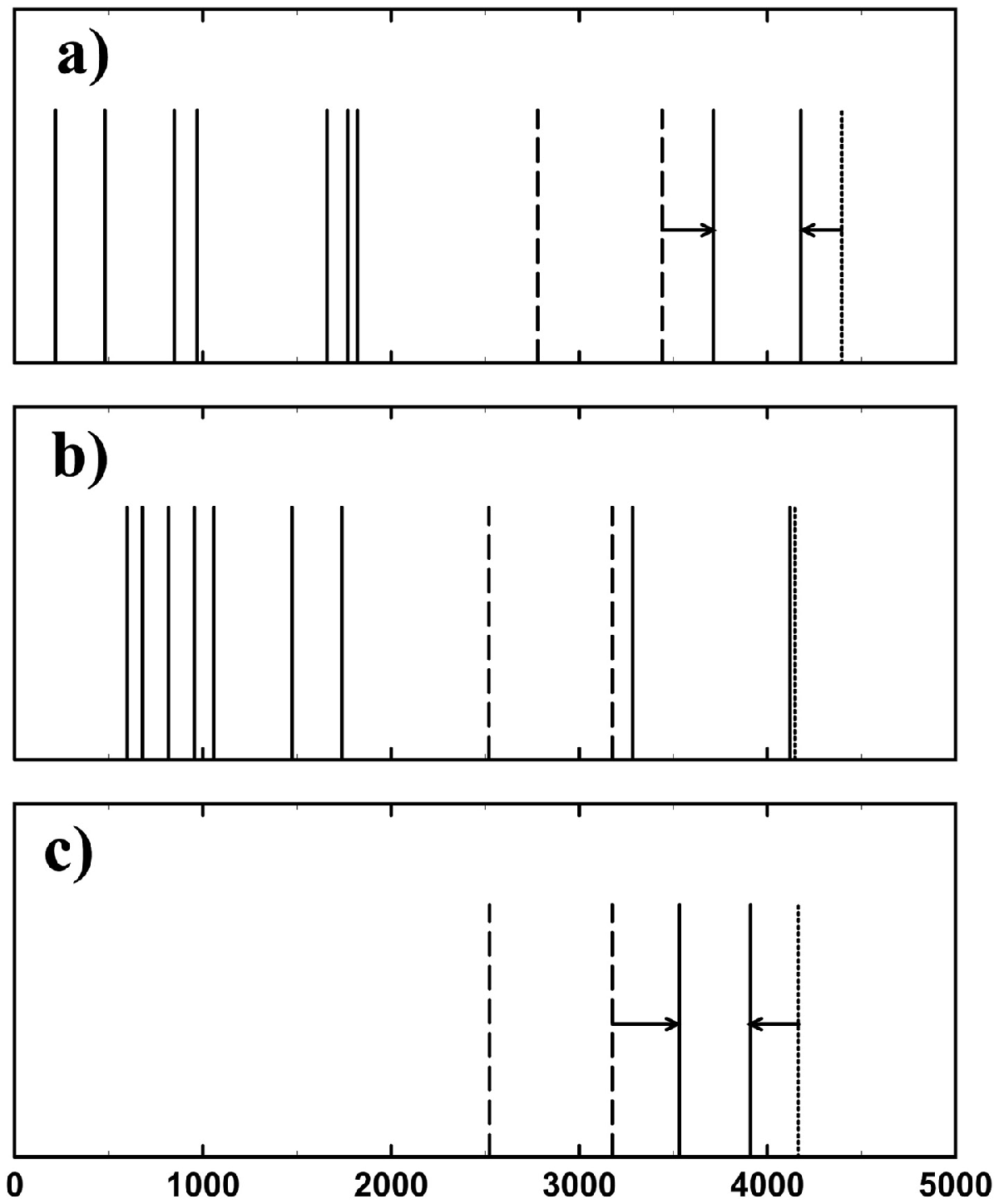}
\caption{Comparison between the frequencies of the H$_{2}$ and H$_{3}^+$ molecules within H$_{5}^+$ cluster with
H$_{2}$ and H$_{3}^+$ isolated molecules. a) Harmonic approximation to PES of ref.\cite{WVO94}. b) CFQMC
computations using the PES of ref.\cite{WVO94}. c) Experiment \cite{OYL88}. Full lines represent the frequencies
of the H$_{5}^+$ cluster. Dotted lines represent the frequency of the isolated H$_{2}$ molecule. Dashed lines
represent the frequencies of the isolated H$_{3}^{+}$ molecule. \label{fig1}}
\end{figure}

\begin{figure}
\includegraphics[0,0][424,330]{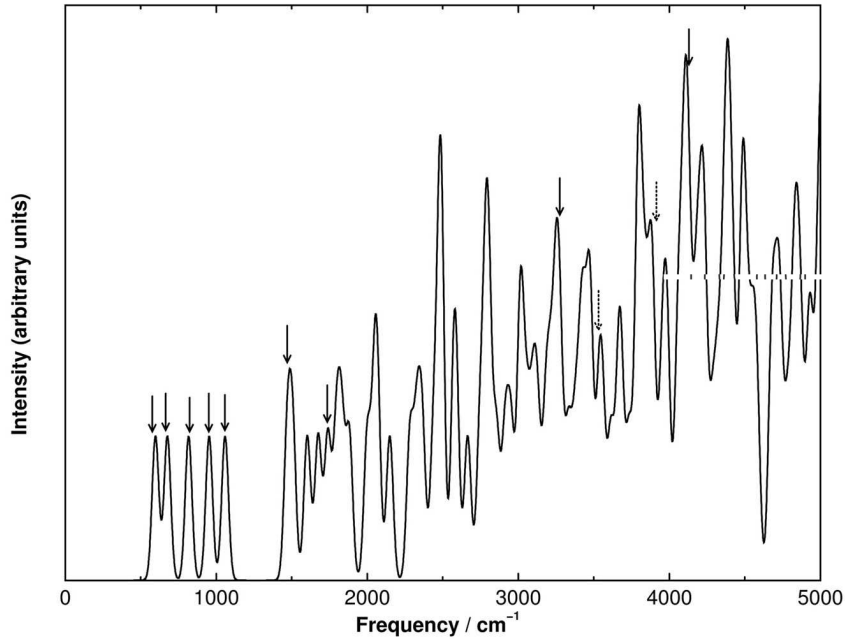}
\caption{ The full CFQMC-computed  vibrational spectrum of the H$_{5}^+$ cluster up to 5000 cm$^{-1}$. The lines
were broadened by 30 cm$^{-1}$. The full arrows correspond to the fundamental frequencies. The dotted arrows
correspond to the measured frequencies of the H$_{2}$ and H$_{3}^+$ in the H$_{5}^+$ cluster\cite{OYL88}.}
\label{fig2}
\end{figure}

\begin{figure}
\includegraphics[0,0][424,424]{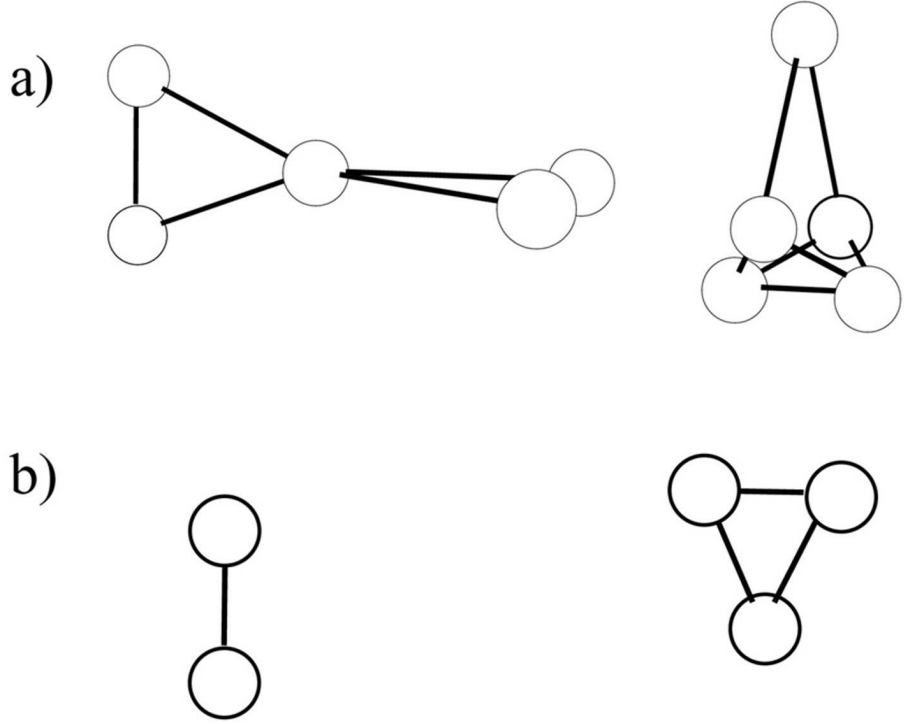}
\caption{ a) Two candidates for global minima of H$_{5}^{+}$ molecules. b) Global minimum of the H$_{5}^{+}$ as
defined by the PES of ref. \cite{WVO94}} \label{fig3}
\end{figure}
\end{center}
%$^{}$\\
%$^{}$\\
%$^{}$\\
%$^{}$\\
%$^{}$\\
%$^{}$\\
%$^{}$\\
%$^{}$\\
%$^{}$\\
%$^{}$\\
%$^{}$\\
%$^{}$\\
%$^{}$\\
%$^{}$\\
%$^{}$\\
%$^{}$\\
%$^{}$\\
\newpage

%\squeezetable
\begin{table}

\caption{
Vibrational frequencies in the harmonic approximation of the
H$_5^{+}$ ({\it{cm}$^{-1}$})
}
\begin{tabular}{cccccc}
\hline
Mode & PES of Ref. \cite{WVO94}$^a$ & MP2$^b$  &
 Prosmiti I $^c$ & Prosmiti II $^d$ &  Main feature \\
\hline
1 ($a_2$) &  215 &  211 &  206   &  206 &  H$_3^+$-H$_2$ torsion\\
2 ($a_1$) &  477 &  482 &  495   &  502 & H$_3^+$-H$_2$ stretch\\
3 ($b_2$) &  848 &  822 &  812   &  815 &  H$_3^+$ rocking\\
4 ($b_1$) &  970 &  881 &  866   &  868 &  Skeletal motion\\
5 ($b_1$) & 1660 & 1201 & 1170   &  1174 & Skeletal motion\\
6 ($a_1$) & 1768 & 1862 & 1838   &  1840 &  H$_3^+$ bending\\
7 ($b_2$) & 1819 & 2186 & 2131   &  2134 &  H$_3^+$ asym. stretch\\
8 ($a_1$) & 3714 & 3761 & 3668   &  3670 &  H$_3^+$ sym. stretch\\
9 ($a_1$) & 4177 & 4247 & 4115   &  4118 &  H$_2$ stretching\\
\hline
\end{tabular}$^{}$\\
$^a$ Present work.\\
$^b$ MP2/6-311G(p,3d), present work.\\
$^c$ CCSD(T)/aug-cc-pVTZ \cite{a1}.\\
$^d$ QCISD(T)/cc-pVQZ Ref. \cite{RAPG01}.
\end{table}

\begin{table}
\caption{{Vibrational frequencies of the H$_5^{+}$}({\it{cm}$^{-1}$})}
\begin{tabular}{ccccc}
\hline
States & CFQMC$^a$ & \v{S}pirko \cite{WVO94} & MP2-VSCF $^a$ & Exp \cite{OYL88} \\
\hline
1 &  598 &  622 &   596  &  - \\
2 &  676 &   -  &   776  &  - \\
3 &  817 &   -  &   879  &  - \\
4 &  952 &  973 &   989  &  - \\
5 & 1058 & 1238 &  1142  &  - \\
6 & 1471 & 1383 &  1173  &  - \\
7 & 1738 &   -  &  1735  &  - \\
8 & 3281 & 3471 &  3515  &  3532 \\
9 & 4117 & 3897 &  3921  &  3910 \\
\hline
\end{tabular}$^{}$\\
$^a$ Present work.
\end{table}

\end{document}